\documentclass[%
 aip,
 pop,%
 amsmath,amssymb,showpacs,
reprint,%
numerical
]{revtex4-1}

\usepackage{graphicx} 
\usepackage{dcolumn}
\usepackage{bm}
\usepackage{braket}
\usepackage{color}
\usepackage[usenames,dvipsnames]{xcolor}
\usepackage{multirow}

\newcommand{\vect}[1]{\boldsymbol{#1}}
\newcommand{\norm}[1]{\lVert#1\rVert}
\newcommand{\refeq}[1]{(\ref{#1})}

\begin{document}

\preprint{AIP/123-QED}

\title[Plasma heating power dissipation in low temperature hydrogen plasmas]{Plasma heating power dissipation in low temperature hydrogen plasmas}

\author{J. Komppula}
\email{jani.komppula@jyu.fi}
 \affiliation{Department of Physics, University of Jyv{\"a}skyl{\"a}, P.O. Box 35, FI-40014 University of Jyvaskyla, Finland}
\author{O. Tarvainen}%
 \affiliation{Department of Physics, University of Jyv{\"a}skyl{\"a}, P.O. Box 35, FI-40014 University of Jyvaskyla, Finland}
\date{\today}

\begin{abstract}
Theoretical framework for power dissipation in low temperature plasmas in corona equilibrium is developed. The framework is based on fundamental conservation laws and reaction cross sections and is only weakly sensitive to plasma parameters, e.g. electron temperature and density. The theory is applied to low temperature atomic and molecular hydrogen laboratory plasmas for which the plasma heating power dissipation to photon emission, ionization and chemical potential is calculated. The calculated photon emission is compared to recent experimental results.
\end{abstract}

\maketitle
\section{Introduction}
Photon emission of low temperature plasmas is relevant for applications of plasma sources, e.g. thin film processing \cite{Tian_2005_controlling_VUV,plasma_ALD_review,Potts_2013_VUV_ALD}, photolithography \cite{plasma_etching_review} and ion sources \cite{Tarvainen2011113,Laulainen_NIBS_2014}. The plasma energy efficiency is critical in plasma thrusters \cite{Gascon_2003} and thermonuclear fusion machines \cite{Wesson_1989_Disruptions_in_JET}, for example. Low temperature plasmas are also found in interstellar space \cite{Shaw_2005_H_2_astrophysics}, where photon emission affects the plasma dynamics \cite{Meerson_1996_Nonlinear_dynamics}.

Currently, there are no practical methods to estimate the relevance of plasma heating power dissipation channels, e.g. photon emission and particle losses, with analytical expressions. The Saha equation and Boltzman equilibrium \cite{DeglInnocenti_2014_book} require assumptions that are unrealistic in typical laboratory plasmas. On the other hand, radiative condensation models describe only nonlinear dynamics of instabilities \cite{Meerson_1996_Nonlinear_dynamics} and excimer light source models cover only the energy transfer of electrons to neutral gas \cite{Salvermoser_2003_xenon_excimer_theory}. Furthermore, simulations about VUV-emission of low temperature hydrogen plasmas are rarely found \cite{Batishchev_2000_helicon_simulation}. We present a mathematical framework (in analytical form) for power dissipation in low temperature plasmas. The theory is based on fundamental conservation laws and cross sections and is only weakly sensitive to plasma parameters. Fractions of plasma heating power dissipation via photon emission, ionization and chemical potential (dissociation and excitation to metastable states including vibrational excitation) are calculated in steady state for low temperature hydrogen plasmas. The photon emission, which is the only (directly) measurable quantity, is compared to experimental results.

\section{Theoretical background}

For the clarity of presentation and due to limited cross section data, atomic and molecular hydrogen plasmas are treated under the following assumptions:  
\begin{enumerate}
\item the plasma is electro-positive consisting of electrons, singly charged ions and neutrals (either atoms or molecules at lowest vibrational level)
\item excited states of particles (neutrals in hydrogen plasmas) are either radiative or metastable. Radiative states decay instantly via photon emission while metastable states have significantly longer lifetimes and can be de-excited non-radiatively.
\item excitations and ionization occur from the ground state of the neutrals
\item the average electron energy is equal or higher than the average neutral and ion energy, i.e. excitation and ionization are caused by electron impact. 
\item radiation is emitted by electronic transitions
\item the plasma is optically thin
\end{enumerate}
These assumptions limit the suitable range of plasma parameters. At least the lowest electronic states of atoms and molecules  must be in corona equilibrium, i.e. the plasma density is \textless $10^{14}$--$10^{16}$~1/cm$^3$ as presented in Refs. \cite{Goto_2002_Ionization_recombination_ratio,komppula_VUV_diagnostics}. The (molecular) ionization degree must be low in order to minimize the influence of other hydrogen species (e.g. H$_2^+$ and H$_3^+$). These limitations are fullfilled by most low temperature laboratory plasmas \cite{Conrads_2000_plasma_sources,Bacal_2015_Negative_hydrogen_ion_production}.

The conservation of the electric charge, mass and energy govern the plasma dynamics, and can be described with continuity equations. The transport of particles is described by
\begin{equation}
\label{eq:continuity_equation}
\frac{\partial n_\alpha}{\partial t}+\nabla\cdot \vect{j}_\alpha =r_\alpha,
\end{equation}
where $n$ is the density, $\vect{j}$ is the particle flux\footnote{In this study vector $\vect{j}_\alpha$ represents the particle flux and, therefore, the electric current is calculated by multplying $j_{\alpha}$ by electron charge.} and $r$ is the local volumetric generation rate of particle species $\alpha$ (electrons $e$, ions $i$ or neutrals $n$). The particle flux can be calculated from the local velocity probability distribution function (VPDF), $f(\vect{v})$, as 
\begin{equation}
\label{eq:particle_flow}
\vect{j}_\alpha =n_\alpha\int_{\vect{v}}\norm{\vect{v}}\,f_\alpha(\vect{v})\,d\vect{v},
\end{equation}
where $\vect{v}$ is the particle velocity

Conservation of charge and mass are connected by ionization and recombination. The reaction rates $r_\alpha$ describe the difference between volumetric ionization $r_\text{inz}$ and recombination $r_\text{rec}$ rates, i.e.
\begin{equation}
\label{eq:reaction rate}
r_e=r_i=-r_n=r_\text{inz}-r_\text{rec}.
\end{equation}

The plasma energy is carried by particles and electromagnetic fields (radiation). The rate of particle energy (enthalpy) transfer (per unit volume) can be expressed as 
\begin{equation}
\label{eq:energy_density_of_particles}
\frac{\partial u_\text{p}}{\partial t}+\nabla\cdot \vect{S}_\text{p}=r_\text{p},
\end{equation}
where $u_\text{p}$ is the energy density of the plasma particles, $\vect{S}_\text{p}$ is the directional particle energy flux density and $r_p$ is the volumetric rate of energy conversion from the electromagnetic field to particle energy. The energy carried by particles consists of kinetic energy and chemical potential. In this study the chemical potential is defined to be the potential energy carried by particles in comparison to a reference particle \footnote{The reference particle is an average neutral particle in the initial state of the closed system or an average incoming neutral particle in the open system.}. In the other words, it consists of ionization potential, dissociation potential and excitation energy of metastable states. Thus, the energy flux density $\vect{S}_\text{p}$  can be divided into kinetic energy $\vect{S}_\text{k}$ and chemical potential $\vect{S}_\mu$, i.e.
\begin{eqnarray}
\label{eq:divided_energy_density_of_particles}
\nonumber \vect{S}_p=&\vect{S}_\text{k}+\vect{S}_\mu\\
=&\sum_\alpha n_\alpha \int_{\vect{v}}\left(\frac{1}{2}m_\alpha\norm{\vect{v}}^2+\braket{\Delta E_{\mu,\alpha}}\right)\norm{\vect{v}} f(\vect{v}) d\vect{v},
\end{eqnarray}
where $m_{\alpha}$ is the mass and $\braket{\Delta E_{\mu,\alpha}}$ is the average chemical potential. 

The volumetric rate of energy transfer of the electromagnetic field is described by Poynting's theorem,
\begin{equation}
\label{eq:poyntings_theorem}
\frac{\partial u_\text{emf}}{\partial t}+\nabla\cdot \vect{S}_\text{emf}=r_\text{emf},
\end{equation}
where $u_\text{emf}$ is the energy density of the electromagnetic field, $\vect{S}_\text{emf}$ is the Poynting vector corresponding to the directional energy flux density of the electromagnetic field and $r_\text{emf}$ describes the volumetric rate of energy conversion from the plasma particles to the electromagnetic field and vice versa, e.g. via individual photon interactions and interaction between the electromagnetic field and charged particles ($\vect{j}\cdot \vect{E}$), i.e. Joule heating. 

Electromagnetic radiation can be categorized based on its origin and type of interaction with the plasma. Micro- and radio-waves interact (mainly) collectively with the plasma and are referred as EM-waves hereafter. Higher energy electromagnetic radiation, referred as photons, interacts with the plasma (mainly) through individual photon--particle  interactions.  The dominating photon--particle processes are excitation and ionization and the opposite reactions of recombination and radiative transitions between electronic quantum states. Thus, the Poynting vector can be divided to two parts describing the EM waves $\vect{S}_\text{em}$ and photons (light) $\vect{S}_\nu$, i.e. $\vect{S}_\text{emf}=\vect{S}_\text{em}+\vect{S}_{\nu}$.

Conservation of energy requires that $r_\text{p}+r_\text{emf}=0$ and, therefore, the total rate of energy transfer in the plasma $\frac{\partial u_\text{tot}}{\partial t}$ (sum of equations \refeq{eq:energy_density_of_particles} and \refeq{eq:poyntings_theorem}) can be expressed as
\begin{equation}
\label{eq:total_energy_density}
-\frac{\partial u_\text{tot}}{\partial t}=\nabla\cdot \vect{S}_\text{k}+\nabla\cdot \vect{S}_\mu+\nabla\cdot \vect{S}_\text{em}+\nabla\cdot \vect{S}_{\nu}.
\end{equation}

Radiative states can be formed by electron impact excitation or by recombination. Thus,
\begin{equation}
\label{eq:photon_emission_power}
\nabla\cdot\vect{S}_{\nu}=\braket{E_{\nu \text{e}}}r_{\nu \text{e}}+\braket{E_{\nu \text{r}}} r_{\text{rec}}+\mathcal{O}_\nu,
\end{equation}
where $\braket{E_{\nu \text{e}}}$ is the average radiative energy of all radiative states, $r_{\nu e }$ is the total excitation rate to radiative states (excluding recombination), $\braket{E_{\nu_ \text{r}}}$ is the average radiative energy emitted subsequent to recombination, $r_{\text{rec}}$ is the total volumetric recombination rate and $\mathcal{O}_\nu$ describes less significant destructive processes of photons such as photo-ionization and -excitation \cite{Beynon_1965_H_Photoionization,Golubovskii_2013_metastable_transport}. 

The chemical potential $\nabla\cdot \vect{S}_\mu$ can be divided into ionization potential and chemical potential associated to dissociation and excitation energies of vibrational levels and metastable electronic states, i.e.
\begin{equation}
\label{eq:metastable_excitation_power}
\nabla\cdot \vect{S}_\mu=E_\text{inz}(r_\text{inz}-r_\text{rec})+\braket{\Delta E_{\mu c}}r_{\mu c}+\mathcal{O}_\mu,
\end{equation}
where $E_\text{inz}$ is the ionization energy, $\braket{\Delta E_{\mu c}}$ is the average change of chemical potential in discrete processes,  $r_{\mu c}$ is the corresponding total volumetric reaction rate and $\mathcal{O}_\mu$ describes less significant destructive processes of chemical potential, e.g. collisional quenching \cite{2S_Collisional_Quenching,H2_c_quenching} or electron impact de-excitation \cite{Sartori_1998_electron_impact_H2_deexcitation_triplet,Golubovskii_2013_metastable_transport} of metastable states.

The kinetic energy of charged particles is typically affected by the change of the energy carried by the electromagnetic waves ($\vect{S}_\text{em}$), the work done by the electric field ($\vect{j}\cdot \vect{E}$) and electron impact processes. On the contrary, the kinetic energy of neutrals is typically  affected only by discrete events, e.g. electron impact excitation to repulsive electronic states. Thus, $\nabla\cdot \vect{S}_\text{k}$ can be divided into kinetic energy of charged particles and kinetic energy of neutrals, 
\begin{equation}
\label{eq:kinetic_power}
\nabla\cdot \vect{S}_\text{k}=\nabla\cdot \vect{S}_\text{kc}+\braket{\Delta E_\text{kn}}r_\text{kn}+\mathcal{O}_\text{kn},
\end{equation}
where $\nabla\cdot \vect{S}_\text{kc}$ describes the change of the kinetic energy carried by charged particles, $\braket{\Delta E_\text{kn}}$ is the average change of the kinetic energy of neutrals in discrete processes, $r_\text{kn}$ is the corresponding total volumetric reaction rate and $\mathcal{O}_\text{kn}$ describes less significant processes, e.g. kinetic energy loss in vibrational excitations.

The average energies ($\braket{E_{\nu \text{e}}}$, $\braket{E_{\nu \text{r}}}$, $\braket{\Delta E_{\mu c}}$ and $\braket{\Delta E_\text{kn}}$ in equations \refeq{eq:photon_emission_power}--\refeq{eq:kinetic_power}) are rate coefficient weighted sums of reaction energies, i.e.
\begin{equation}
\label{eq:average_energy_general}
\braket{E_\alpha}=\frac{\sum_i E_{\alpha,i}\braket{\sigma_{\alpha,i}v}}{\sum_i\braket{\sigma_{\alpha,i}v}},
\end{equation}
where $E_{\alpha,i}$ is the energy and $\braket{\sigma_{\alpha,i}v}$ the rate coefficient of reaction $i$. 

Ionization is the primary process for sustaining the plasma. Therefore, it is rational to connect the volumetric process rates to the volumetric ionization rate. Different rate coefficients $r_\alpha$ can be expressed with the ionization rate $r_\text{inz}$ and the ratio of the reaction rates $k_{\alpha}$, i.e. 
\begin{equation}
\label{eq:ratio_of_rates_general2}
r_\alpha=k_{\alpha} r_\text{inz}=\frac{r_\alpha}{r_\text{inz}} r_\text{inz}.
\end{equation}
Because it is assumed that excitations occur only from the ground state of neutrals by electron impact, Eq.  \refeq{eq:ratio_of_rates_general} does not depend on the electron or neutral particle densities and it can be expressed with rate coefficients as

\begin{equation}
\label{eq:ratio_of_rates_general}
r_\alpha=k_{\alpha} r_\text{inz}=\frac{\sum_i \braket{\sigma_{\alpha,i}v}}{\braket{\sigma_\text{inz}v}} r_\text{inz},
\end{equation}
where the subscript $i$ refers to different branches of the excitation process.

\section{Energy balance of laboratory plasmas}
A laboratory plasma can be described as an open thermodynamic system, which exchanges energy, matter and electric charge with its surroundings. Equations \refeq{eq:continuity_equation}--\refeq{eq:kinetic_power} describe the local (thermodynamic) energy balance. The system, however, can be studied analytically only globally, i.e. applying divergence theorem to Eq. \refeq{eq:total_energy_density}. This yields
\begin{eqnarray}
\label{eq:total_energy}
\nonumber-\frac{\partial U_\text{tot}}{\partial t}=&\int_V \left( \nabla\cdot \vect{S}_\text{k}+\nabla\cdot \vect{S}_\mu +\nabla\cdot \vect{S}_\text{em}+\nabla\cdot \vect{S}_{\nu}\right ) dV\\ 
&=\int_A\left(  \vect{S}_\text{k}+\vect{S}_\mu+\vect{S}_\text{em}+\vect{S}_{\nu}\right )\cdot d\vect{A},
\end{eqnarray}
where $U_{tot}$ is the total energy of the system (plasma particles and EM-fields) and $V$ is the volume enclosed by a surface with an area $A$. The volumetric integrals in Eq. \refeq{eq:total_energy} correspond to the energy conversion within the volume and surface integrals correspond to energy flow through the closed surface. 

Thus, Eq. \refeq{eq:total_energy} can be rewritten as 
\begin{eqnarray}
\label{eq:total_power}
-\frac{\partial U_\text{tot}}{\partial t}&=P_{\Delta\text{k}}+P_{\Delta\mu}+P_{\Delta\text{em}}+P_{\Delta\nu}\\
\nonumber&=P_{\text{J,k}}+P_{\text{J,}\mu}+P_{\text{J,em}}+P_{\text{J,}\nu},
\end{eqnarray}
where the subscript $\Delta$ represents the power conversion in the volume and subscript $\text{J}$ the total power flow through the surface. 

Substituting Eqs. \refeq{eq:average_energy_general} and \refeq{eq:ratio_of_rates_general} into volume integrals of Eqs. \refeq{eq:photon_emission_power}, \refeq{eq:metastable_excitation_power} and \refeq{eq:kinetic_power} yields
\begin{eqnarray}
&P_{\Delta\text{k}}=P_{\Delta\text{kc}}+k_{\text{kn}}\braket{\Delta E_\text{kn}}R_\text{inz}+\mathcal{O}_\text{kn},\label{eq:delta_k}\\
&P_{\Delta\mu}=E_\text{inz}(R_\text{inz}-R_\text{rec})+k_{\mu c}\braket{\Delta E_{\mu c}}R_\text{inz}+\mathcal{O}_\mu\ \ \ \ \ \label{eq:delta_exc}\\
&P_{\Delta\nu}=k_{\nu\text{e}}\braket{E_{\nu\text{e}}}R_\text{inz}+\braket{E_{\nu\text{r}}} R_\text{rec}+\mathcal{O}_\nu \label{eq:delta_nu}
\end{eqnarray}
where $R_\alpha$ is the total rate of a given process in the volume, $\braket{E_\alpha}$ is the associated average energy and $k_\alpha$ the average ratio of the rate coefficients (corresponding reaction vs. ionization, Eq. \refeq{eq:ratio_of_rates_general}).

The system exchanges kinetic energy with the surroundings via escaping electron-ion pairs $P_{\text{J,kp}}$ (carrying an average kinetic energy of $\braket{\Delta E_{p}}$), incoming particles dissipating energy through collisions ($P_{\text{J,kH}}$) and escaping neutral particles ($P_{\text{J,kn}}$), i.e.
\begin{eqnarray}
\label{eq:total_kinetic_power_simplified_2}
&P_\text{J,k}=P_{\text{J,kp}}+P_{\text{J,kH}}+P_{\text{J,kn}}\\
\nonumber=&(R_\text{inz}-R_\text{rec})\braket{\Delta E_{p}}+P_{\text{J,kH}}+P_{\text{J,kn}}.
\end{eqnarray}

Chemical potential flow $P_{\text{J,}\mu}$ can be divided to ionization potential, $P_{\text{J,inz}}$, and chemical potential associated to dissociation and excitation energies, $P_{\text{J,}\mu c}$, as in Eq. \refeq{eq:metastable_excitation_power}, i.e.
\begin{eqnarray}
\label{eq:total_kinetic_power_simplified}
&P_{\text{J,}\mu}=P_{\text{J,inz}}+P_{\text{J,}\mu c}
\end{eqnarray}

Power dissipation through photon emission, ionization and chemical potential can be studied analytically using Eqs. \refeq{eq:total_power}--\refeq{eq:total_kinetic_power_simplified} if $\partial U_\text{tot}/\partial t$ is known. In time invariant steady state $\partial U_\text{tot}/\partial t$=0, which means that the power used for sustaining the plasma is converted from one form to another inside the system, i.e. $\sum_\alpha P_{\Delta\alpha}$=0 and $\sum_\alpha P_{J,\alpha}$=0 (Eq. \refeq{eq:total_power}). The plasma is typically sustained by the power carried by EM waves and incoming particles ($P_{\text{J,kH}}+P_\text{J,em}$) which must correspond to photon emission power and power (kinetic + excitation) carried by escaping particles, i.e.
\begin{equation}
\label{eq:power_balance}
P_\text{heat}=P_{\text{J,kH}}+P_\text{J,em}=P_{J,\nu}+P_{\text{J,}\mu}+P_{\text{J,kp}}+P_{\text{J,kn}}.
\end{equation}
Therefore, the ratios of the total photon emission, ionization and chemical potential powers to the plasma heating power can be expressed as
\begin{eqnarray}
\frac{P_{J,\nu}}{P_\text{heat}}=\frac{P_{J,\nu}}{P_{J,\nu}+P_{\text{J,}\mu}+P_{\text{J,kp}}+P_{\text{J,kn}}}\label{eq:fraction_of_photon_emission}\\
\frac{P_{\text{J,inz}}}{P_\text{heat}}=\frac{P_{\text{J,inz}}}{P_{J,\nu}+P_{\text{J,}\mu}+P_{\text{J,kp}}+P_{\text{J,kn}}}\label{eq:fraction_of_ionization}\\
\frac{P_{\text{J,}\mu c}}{P_\text{heat}}=\frac{P_{\text{J,}\mu c}}{P_{J,\nu}+P_{\text{J,}\mu}+P_{\text{J,kp}}+P_{\text{J,kn}}}\label{eq:fraction_of_enthalphy}
\end{eqnarray}

Power conversion to photon emission, ionization and chemical potential as well as change of the kinetic energy of neutrals can be calculated from Eqs. \refeq{eq:delta_nu}--\refeq{eq:delta_k}. The destructive terms $\mathcal{O}_\alpha$ depend strongly on the plasma properties. In optically thin plasma destructive processes of photons are negligible, i.e. $\mathcal{O}_\nu$=0 . In typical hydrogen discharges the majority of metastable states escape from the plasma \cite{Bonnie_1988_metastable_H2_in_multicusp_ion_source}. Their most significant de-excitation processes are radiative decay induced by collisions and ionization \cite{Bonnie_1988_metastable_H2_in_multicusp_ion_source,Sartori_1998_electron_impact_H2_deexcitation_triplet,Glass-Maujean_1989_2S_Quenching}. Thus, $\mathcal{O}_\mu$=0 corresponds to minimum photon emission, minimum ionization and maximum chemical potential. Moreover, it can be argued that inelastic neutral collisions do not affect the energy conversion, i.e. $\mathcal{O}_\text{kn}$=0. Because the destructive processes are negligible, the energy conversion terms ($P_{\Delta\nu}$ and $P_{\Delta\mu}$) correspond to escaping powers ($P_{J,\nu}$ and $P_{\text{J,}\mu}$). Furthermore, $P_{\text{J,kn}}$ corresponds to $k_{\text{kn}}\braket{\Delta E_\text{kn}}R_\text{inz}$. Therefore, substituting equations \refeq{eq:delta_k}--\refeq{eq:total_kinetic_power_simplified} into equations \refeq{eq:fraction_of_photon_emission}--\refeq{eq:fraction_of_enthalphy} yields
\begin{eqnarray}
\label{eq:fraction_of_photon_emission_open}
\frac{P_{J,\nu}}{P_\text{heat}}&\geq\left (k_{\nu\text{e}}\braket{E_{\nu\text{e}}}+\braket{E_{\nu\text{r}}}\frac{R_\text{rec}}{R_\text{inz}}\right)\\
\nonumber&/  \Big ( k_{\nu\text{e}}\braket{E_{\nu\text{e}}}+k_{\mu c}\braket{\Delta E_{\mu c}}+k_{\text{kn}}\braket{\Delta E_\text{kn}}+E_\text{inz}\\
\nonumber&+\braket{\Delta E_{p}}+(\braket{E_{\nu\text{r}}}-E_\text{inz}-\braket{\Delta E_{p}})\frac{R_\text{rec}}{R_\text{inz}} \Big ).
\end{eqnarray}

\begin{eqnarray}
\label{eq:fraction_of_ionization_open}
\frac{P_{\text{J,inz}}}{P_\text{heat}}&\geq\left (E_\text{inz}(1-\frac{R_\text{rec}}{R_\text{inz}})\right)\\
\nonumber&/  \Big ( k_{\nu\text{e}}\braket{E_{\nu\text{e}}}+k_{\mu c}\braket{\Delta E_{\mu c}}+k_{\text{kn}}\braket{\Delta E_\text{kn}}+E_\text{inz}\\
\nonumber&+\braket{\Delta E_{p}}+(\braket{E_{\nu\text{r}}}-E_\text{inz}-\braket{\Delta E_{p}})\frac{R_\text{rec}}{R_\text{inz}} \Big ).
\end{eqnarray}

\begin{eqnarray}
\label{eq:fraction_of_enthalphy_open}
\frac{P_{\text{J,}\mu c}}{P_\text{heat}}&\leq\left (k_{\mu c}\braket{\Delta E_{\mu c}}\right)\\
\nonumber&/  \Big ( k_{\nu\text{e}}\braket{E_{\nu\text{e}}}+k_{\mu c}\braket{\Delta E_{\mu c}}+k_{\text{kn}}\braket{\Delta E_\text{kn}}+E_\text{inz}\\
\nonumber&+\braket{\Delta E_{p}}+(\braket{E_{\nu\text{r}}}-E_\text{inz}-\braket{\Delta E_{p}})\frac{R_\text{rec}}{R_\text{inz}} \Big ).
\end{eqnarray}

The results of Eqs. \refeq{eq:fraction_of_photon_emission_open}--\refeq{eq:fraction_of_enthalphy_open} depend only on the properties of the plasma gas through cross sections and two plasma parameters: the average energy of escaping electron-ion pair and the average electron energy. The average electron energy determines the (local) ratio of recombination to ionization \citep{Goto_2002_Ionization_recombination_ratio} ($R_\text{rec}/R_\text{inz}$), ratios of rate coefficients ($k_{\nu\text{e}}$, $k_{\mu c}$ and $k_{\text{kn}}$) and average energies $\braket{E_{\nu\text{e}}}$, $\braket{E_{\nu\text{r}}}$, $\braket{\Delta E_{\mu c}}$ and $\braket{\Delta E_\text{kn}}$. The values of these coefficients vary spatially and their global averages are not linearly proportional to the global average of electron energy.

\section{Low temperature hydrogen plasmas}

Different terms in Eqs. \refeq{eq:fraction_of_photon_emission_open}-- \refeq{eq:fraction_of_enthalphy_open} can be calculated by studying excitation processes of the hydrogen atom and molecule. Excitation cross section data is only available for a limited number of electronic states, which, however, cover the majority of total excitations. Electron impact excitation from the ground state (1S) to electronic states of hydrogen atom up to main quantum number $n=5$ have been taken into account hereafter. Those states decay radiatively to the ground state or metastable  $2S$ state as determined by selection rules and transition probabilities \cite{Wiese_2009_hydrogen_atomic_transitions_probabilities}.

For hydrogen molecule excitation cross section data is mainly available for transitions from the lowest vibrational level of the ground state \cite{Celiberto_2001_cross_section_functions,Janev_2003}, $X^1\Sigma_g^+(\nu$=$0)$. Total electron impact excitation from the ground state to singlet states $B^1\Sigma_u^+$, $C^1\Pi_u$, $D^1\Pi_u$, $EF^1\Sigma_g^+$, $B'^1\Sigma_u^+$, $D'^1\Pi_u$  have been taken into account hereafter. These excitations dominate the total excitation cross section to singlet states and lead to ground state via optically allowed transitions. Vibrational level of the molecule changes in electronic excitation (Franck-Condon principle \cite{Fantz_2006_Franck_condon}) and, therefore, 2.3--3.1~eV of the electronic excitation energy is converted to vibrational excitation (chemical potential) \footnote{Calculated assuming that vibrational transitions follow Franck-Condon factors from Ref. \cite{Fantz_2006_Franck_condon}.}. Direct vibrational electron impact excitation from $X^1\Sigma_g^+(\nu$=$0)$ to $X^1\Sigma_g^+(\nu$=$1)$ is included in the calculations. In the case of homonuclear molecules (such as H$_2$) the vibrational levels are metastable states \cite{Fantz_2006_Franck_condon}. 

Electron impact excitations from the ground state $X^1\Sigma_g^+(\nu$=$0)$ to triplet states $a^3\Sigma^+_g$, $b^3\Sigma^+_u$,  $c^3\Pi_u$ $e^3\Sigma^+_u$ and $d^3\Pi^+_u$ have been taken into account due to available cross section data \cite{Janev_2003}. The lowest triplet state ($b^3\Sigma^+_u$) is repulsive leading to dissociation of the molecule. The resulting atoms carry on average 1.5~eV of kinetic energy \cite{Xiao_2001_H2_dissociation_in_low_temperature_plasma} and the molecular binding energy of 4.52~eV is transferred to chemical potential. The $c^3\Pi_u$ triplet state is metastable. Other triplet states ($a^3\Sigma^+_g$, $e^3\Sigma^+_u$, $d^3\Pi^+_u$) populate the repulsive $b^3\Sigma^+_u$ state via radiative transitions.

\begin{figure}
 \begin{center}
 \includegraphics[width=0.45\textwidth]{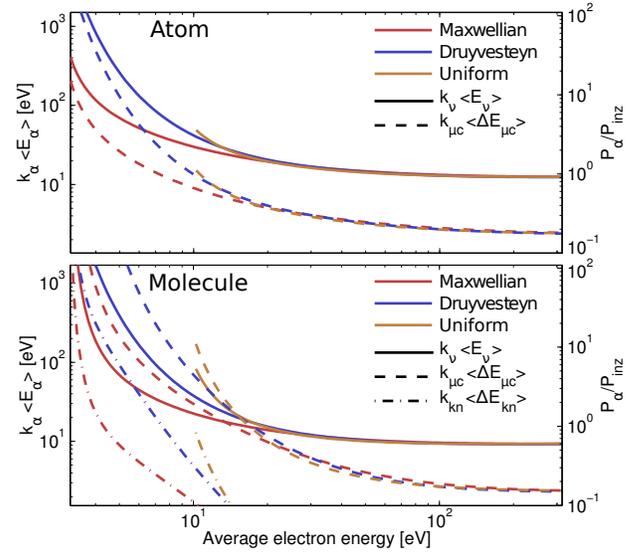}
 \caption{\label{fig:coefficients} Coefficients $k_{\nu \text{e}}\braket{E_{\nu \text{e}}}$, $k_{\mu c}\braket{E_{\mu}}$ and $k_{\text{kin}}\braket{E_{\text{kin}}}$ (Eqs. \refeq{eq:delta_nu}--\refeq{eq:delta_k}) as a function of the average electron energy of different VPDFs. The secondary vertical axis is the ratio of volumetric powers of a particular process to ionization. The cross section data is from Ref.\cite{Janev_2003}.}%
 \end{center}
 \end{figure}

The electron VPDF of low temperature plasmas is typically non-Maxwellian \cite{Lieberman_2005_book}. Inelastic electron impact processes are most sensitive to the high energy tail of the VPDF ($E_e$\textgreater8~eV) due to threshold energies of the electronic excitations and ionization. Mathematical extremes of the VPDF can be described with Maxwellian \cite{Lieberman_2005_book,Behringer_1994_EEDF_distributions}, Druyvesteyn-like \cite{Lieberman_2005_book,Behringer_1994_EEDF_distributions} or uniform distribution \cite{Graham_1995_kinetics_of_negative_hydrogen_ions}. These three distributions\footnote{Maxwellian and Druyvesteyn distributions are calculated with equation 5 in Ref. \cite{Behringer_1994_EEDF_distributions} by using parameters $\nu$=1 and $\nu$=2, respectively. Uniform distribution is calculated with equation 3 in Ref. \cite{komppula_VUV_diagnostics}. The ionization potential limits the minimum average energy of the uniform distribution.} have been used for calculating $k_{\nu \text{e}}\braket{E_{\nu \text{e}}}$, $k_{\mu c}\braket{\Delta E_{\mu c}}$ and $k_{\text{kn}}\braket{\Delta E_{\text{kn}}}$ (Eqs. \refeq{eq:delta_k}--\refeq{eq:delta_nu}), presented in Fig. \ref{fig:coefficients} for atomic and molecular hydrogen as a function of the average electron energy. 

\begin{figure}
 \begin{center}
 \includegraphics[width=0.45\textwidth]{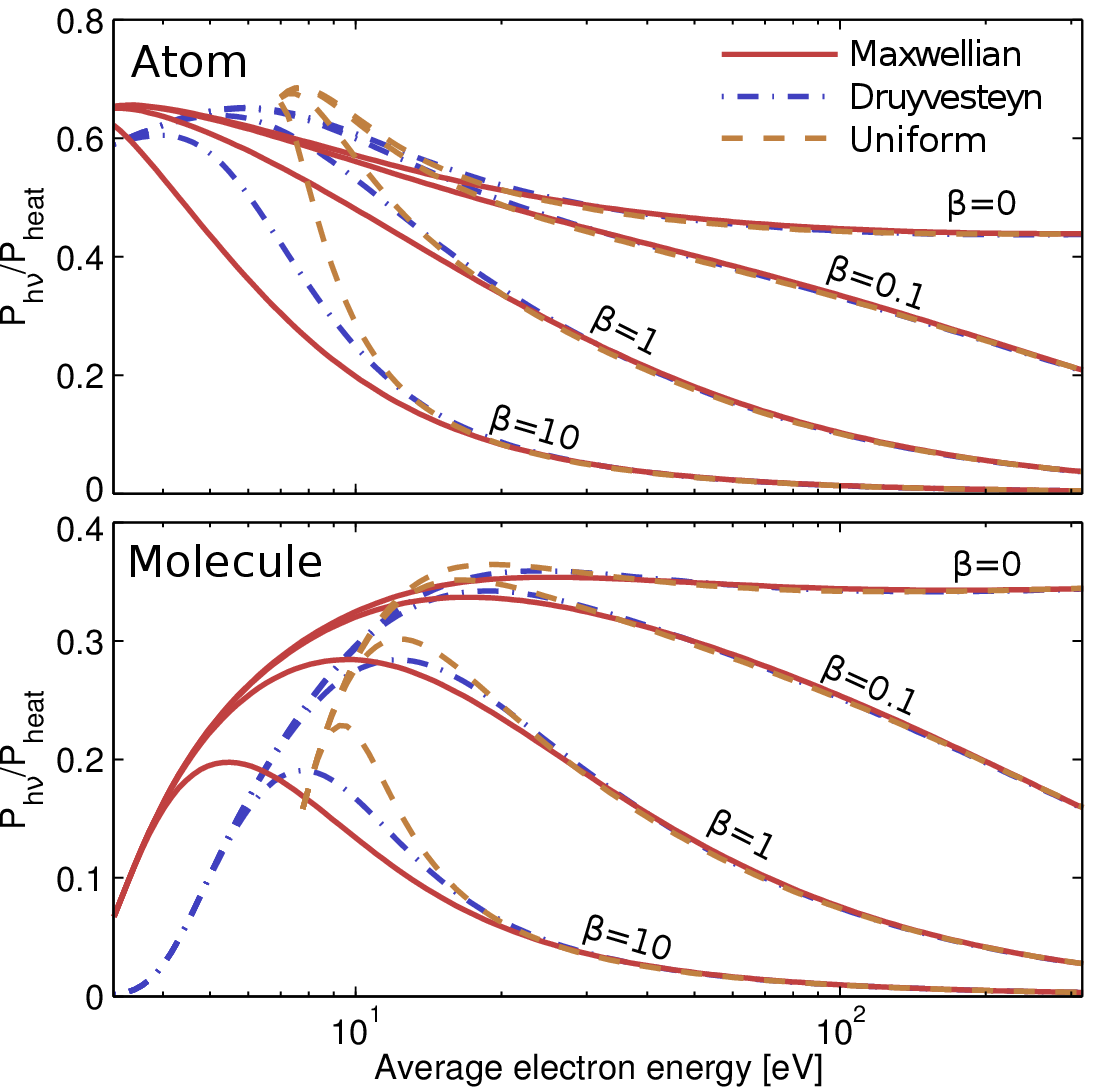}
 \caption{\label{fig:minimum_photon_emission}Minimum fraction of power dissipated by photon emission as a function of the average electron energy for different VPDFs. The parameter $\beta$ corresponds to the ratio of the escaping electron-ion pair energy to the average electron energy in the plasma.}%
 \end{center}
 \end{figure}
 
\begin{figure}
 \begin{center}
 \includegraphics[width=0.45\textwidth]{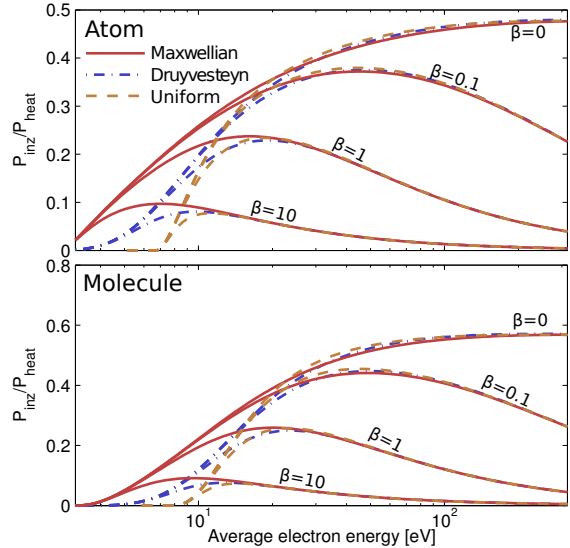}
 \caption{\label{fig:ionization}Minimum fraction of power dissipated in ionization as a function of the average electron energy for different VPDFs. The parameter $\beta$ corresponds to the ratio of the escaping electron-ion pair energy to the average electron energy in the plasma.}%
 \end{center}
 \end{figure}

\begin{figure}
 \begin{center}
 \includegraphics[width=0.45\textwidth]{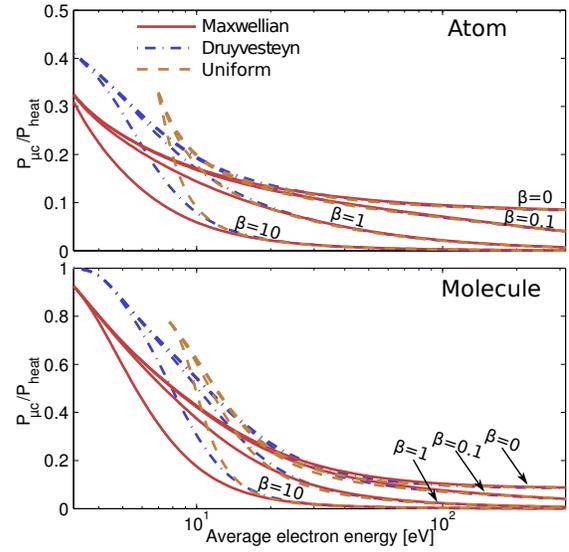}
 \caption{\label{fig:enthalpy}Maximum fraction of power converted to chemical potential associated to dissociation and metastable excitation as a function of the average electron energy for different VPDFs. The parameter $\beta$ corresponds to the ratio of the escaping electron-ion pair energy to the average electron energy in the plasma.}%
 \end{center}
 \end{figure}

The ratio of the average energies of escaping electron-ion pair and the kinetic energy of plasma electrons $\beta$=$\braket{\Delta E_{p}}/\braket{E_e}$ can be used as a parameter to describe the minimum photon emission. The plasma potential does not directly affect $\beta$, because its effect on electron and ion kinetic energies is opposite. However, it effectively increases $\beta$ by determining the minimum energy of escaping electrons. It can be assumed that plasma is ionizing, i.e. recombination rate is negligible in comparison to ionization rate $R_\text{rec}/R_\text{inz}$=0. This is valid in typical laboratory plasmas \citep{Goto_2002_Ionization_recombination_ratio,Bacal_2005_Volume_production}. Because in recombination the electron energy is transferred dominantly to photon emission, the condition $R_\text{rec}/R_\text{inz}$=0 corresponds to minimum photon emission power and maximum ionization power. The ratios $P_{\nu}/P_\text{heat}$, $P_{inz}/P_\text{heat}$ and $P_{\mu c}/P_\text{heat}$ are presented in Figs. \ref{fig:minimum_photon_emission}--\ref{fig:enthalpy} as a function of the average electron energy and parameter $\beta$. The results are calculated using Eqs. \refeq{eq:fraction_of_photon_emission_open}--\refeq{eq:fraction_of_enthalphy_open} and the coefficients presented in Fig. \ref{fig:coefficients}.

The relative importance of different power dissipation channels depends strongly on the average electron energy. At low electron energies ($\braket{E_e}$\textless10~eV) the majority of electron impact processes lead to excitations resulting strong to power dissipation via photon emission and chemical potential (metastable states). If $\braket{E_e}$\textless 5 eV in molecular hydrogen plasma, the majority of the heating power is dissipated without photon emission via molecule dissociation and vibrational excitation. This is because the cross sections of triplet excitations and direct vibrational excitation of molecules are peaked at low electron energies. Ionization has the highest threshold energy and, therefore, its contribution saturates when the average electron energy exceeds the ionization threshold energy.

\section{Conclusions}
Neutral gas in laboratory hydrogen plasmas consists of both atoms and molecules. In terms of heating power dissipation the energy exchange between atoms and molecules is negligible. This means that pure atomic or molecular plasmas are extreme cases. Therefore, the common properties of these two extreme examples can be considered basic properties of hydrogen plasmas.

It can be justified that the VPDFs used here represent the mathematical functional variation of VPDFs found in laboratory plasmas. Therefore, the fractions of plasma heating power dissipation in any (infinitesimal) volume element of the plasma can be described by the data curves in Figs. \ref{fig:minimum_photon_emission}--\ref{fig:enthalpy}. The relevance (fraction) of each power dissipation channel vary spatially in the finite plasma volume due to spatially varying VPDF. However, because the variation of the power dissipation fraction is represented by continuous functions, the mean value theorems for integrals govern that volumetric average of the values is found at certain local values. This means that the fractions of plasma heating power dissipation in specific points (corresponding to a certain VPDF) can be used to describe the fractions of plasma heating power dissipation in the entire plasma volume\footnote{The specific point is typically found in the hot (central) region of the plasma source, where most of the inelastic electron collisions occur.}. 

Altogether this allows making some concluding remarks from Figs. \ref{fig:minimum_photon_emission}--\ref{fig:enthalpy} as basic properties of low temperature hydrogen plasmas:
\begin{enumerate}
\item At least 10\% of plasma heating power is dissipated via photon emission when at least 1\% of the heating power is dissipated in ionization ($P_\text{inz}/P_\text{heat}$\textgreater 0.01). The result is independent of the plasma confinement ($\beta$) at low average electron energies ($\braket{E_e}$\textless20~eV). 

\item At least 15--70\% of heating power is dissipated via photon emission, when at least 10\% of heating power is dissipated in ionization.

\item The efficiency of plasma heating can be described as a fraction of power used for ionization, $P_\text{inz}/P_\text{heat}$ (Fig. \ref{fig:ionization}). The maximum of plasma heating efficiency is achieved when $\braket{E_e}\geq$ 10~eV) independent of the plasma confinement ($\beta$) 

\item The chemical potential associated to dissociation and excitation to metastable states decreases as a function of the average electron energy. At low average electron energies ($\braket{E_e}\leq$ 5~eV) the majority of the heating power is converted to chemical potential in molecular plasmas. This is mostly due to molecule dissociation through excitation to the lowest (repulsive) triplet state and direct vibrational excitation.

\item The relevance of each power dissipation channel is independent of the functional shape of the EEDF, if the average electron energy exceeds 20~eV
\end{enumerate}

In an ideal plasma source (good plasma confinement) the kinetic energy of the electron-ion pair is minimal in comparison to the average electron energy, i.e. $\beta$=0. In other words the electron energy is efficiently dissipated in inelastic collisions before they escape the confinement. This situation is typical in filament driven multicusp arc discharges \cite{Lieberman_2005_book}  which allows some additional conclusions to be made:
\begin{enumerate}
\item Plasma heating efficiency, $P_\text{inz}/P_\text{heat}$, increases as a function of the average electron energy and the maximum is limited to 45--60\%
\item At least 30\% of the heating power dissipates via photon emission, when the plasma heating efficiency  is high ($P_\text{inz}/P_\text{heat}$\textgreater 0.4)
\item The plasma heating efficiency saturates when $\braket{E_e}$\textgreater 50~eV
\end{enumerate}

There are limited number of experiments that can be compared to the presented theory. The only directly measurable quantity is the total photon emission of the plasma. Total photon emission of  at least 15--30\% of the discharge power in VUV-range has been measured with a filament driven arc discharge \cite{komppula_NIBS_2012}. For a 2.45 GHz microwave discharge the corresponding number is approximately 10\% \cite{komppula_2015_2.45_GHz_VUV_power}. These values are consistent with this study. It must be emphasized that it is challenging to determine the total microwave or radio frequency power absorbed by the plasma as discussed in Ref. \cite{komppula_2015_2.45_GHz_VUV_power}.  Batishchev et al have simulated that approximately 25\% of the power of a helicon discharge  dissipates via photon emission \cite{Batishchev_2000_helicon_simulation}, which is also consistent with the results. Furthermore, highest performance of filament driven multicusp arc discharges is achieved when the arc voltage exceeds 100~V ($\braket{E_e}$\textgreater 50~eV) \cite{Bacal_2005_Volume_production,Moehs_2005_Negative_hydrogen_ion_sources}, which indirectly supports the theory. 

More precise verification of the theory would require simultaneous (spatial) measurements of absolute light emission and VPDF. This is not trivial due to challenges of accurate VPDF measurements at high electron energies. Another method could be to compare the measured plasma densities of different elements (e.g. noble gases) to the predicted ionization efficiency given by theory as a function of the arc voltage in the filament driven multicusp discharge. 

The presented theory takes into account the same inelastic electron impact processes than typical plasma simulation codes \cite{Batishchev_2000_helicon_simulation,Capitelli_2006_simulation}. The most significant uncertainties of the theory are related to possible plasma--wall interactions, e.g. the balance between vibrational excitation and de-excitation on the hot surfaces \cite{Bacal_2015_Negative_hydrogen_ion_production}, association of hydrogen atoms \cite{plasma_ALD_review} or self-amplification of electrons \cite{Campanell_2015_Self_amplification_of_electrons}.

\begin{acknowledgments}
This work has been supported by the Academy of Finland under the Finnish Centre of Excellence Programme 2012--2017 (Nuclear and Accelerator Based Physics Research at JYFL)
\end{acknowledgments}
\nocite{*}
\bibliography{powerdissipation}

\end{document}